\documentstyle[11pt,amssymb]{article}

\def\be{\begin{equation}}
\def\ee{\end{equation}}

\title{{\hfill\small\tt Commun.\,Math.\,Phys.\,{\bf 26},\,222-236\,(1972)}\\
{\hfill}\\
Coherent States for Arbitrary Lie Group}
\author{A.M. Perelomov\\
{\small\em Institute for Theoretical and
Experimental Physics, 117259 Moscow, USSR}}
 
\date{}
\begin{document}
\maketitle{}

\begin{abstract}\noindent
The concept of coherent states originally closely related to the nilpotent 
group of Weyl is generalized to arbitrary Lie group. For the simplest Lie 
groups the system of coherent states is constructed and its features are 
investigated.
\end{abstract}

\section{Introduction}
\setcounter{equation}{0} 
In a number of fields of quantum theory, and especially in quantum optics and 
radiophysics, it is convenient to use the system of so called coherent states 
[1--3].

These states are in close connection with the nilpotent group first 
considered by Weyl [4].

A question arises: are there exist analogous systems of states for other Lie 
groups?

The recent paper [5] generalizes the concept of coherent states to some Lie 
groups. However, the method proposed in this paper cannot be applied to all 
Lie groups and, in particular, it is inapplicable to compact groups. Besides, 
with this approach the set of coherent states is noninvariant relative to the 
action of the group representation operators.

The present paper proposes another method to extend the concept of coherent 
state\footnote{\,\,Note in the connection that although some states of such 
type were considered previously, the properties of the system of states as 
a whole do not appear to have been investigated (except for the Weyl 
group).}. This method can be applied to any Lie group and is consistent with 
the action of the group on the set of coherent states (see Section 2). 
Sections 3--5 of the paper deal with construction of the system of 
coherent states and with the investigation of its features for the simplest 
Lie groups.

\section{General Properties of Coherent States}
\setcounter{equation}{0}

Let $G$ be an arbitrary Lie group and $T$ be its irreducible unitary 
representation acting in the Hilbert space ${\cal H}$. A vector of this space 
is denoted by the symbol $|\psi \rangle $, the scalar product of the vector 
$|\psi \rangle$ and $|\varphi \rangle $, linear on $|\psi \rangle $ and 
antilinear on $|\varphi \rangle $, by the symbol $\langle\varphi |\psi 
\rangle $, and the projection operator on the vector 
$|\psi \rangle $ by $|\psi \rangle\langle\psi |$.

Let $|\psi _0\rangle $ be some fixed vector in the space ${\cal H}$. Consider 
the set of vectors $\{|\psi _g\rangle\}$, where $|\psi _g\rangle=T(g)\,
|\psi _0\rangle $ and $g$ goes over all the group $G$. It is easy to see 
that two vectors $|\psi _{g_1}\rangle $ and $|\psi _{g_2}\rangle $ differ 
from one another only by a phase factor $(|\psi _{g_1}\rangle=
e^{i\alpha }\,|\psi _{g_2}\rangle ,\,\,\,|e^{i\alpha }|=1)$, or in other 
words determine the same state only if 
$T\left( g_2^{-1}g_1\right) |\psi _0\rangle =e^{i\alpha }\,|\psi _0\rangle $.

Let $H=\{h\}$ be the set of elements of the group $G$ such that $T(h)\,
|\psi _0\rangle =e^{i\alpha (h)}\,|\psi _0\rangle $. It is evident that 
$H$ is a subgroup of the group $G$ and we denote it as the stationary group 
of the state $|\psi _0\rangle $.

From this construction we see that the vectors $|\psi _g\rangle $ for all $g$ 
which belong to one left coset $G$ on $H$ differ from one another only by a 
phase factor and that these vectors determine the same state.

Selecting in each coset $x$ one represenatative $g(x)$ of the group $G$\,
\footnote{\,\,The group $G$ can be considered as a fiber bundle with the 
base  $M=G/H$, and the fiber $H$. Then the choice of $g(x)$ is the choice 
of a certain cross section of this fiber bundle.}  
we get the set of states $\{|\psi _{g(x)}\rangle\}$, or in abridged form of 
writing $\{|x\rangle\}$ where $|x\rangle\in {\cal H}$, $x\in M=G/H$.

Now we may give the definition of generalized coherent states.

{\em The system of coherent states of the type} $(T,\,|\psi _0\rangle )$ ($T$ 
{\em is the representation of the group} $G$ {\em acting in the some space} 
${\cal H}$ {\em and} $|\psi _0\rangle $ {\em is a fixed vector of this space) 
is called a set of states} $\{|\psi _g\rangle\}$, $|\psi _g\rangle =T(g)\,
|\psi _0\rangle $, 
{\em where $g$ runs over all group} $G$. {\em Let} $H$ {\em be a stationary 
subgroup of the state} $|\psi _0\rangle $. {\em Then the coherent state} 
$|\psi _g\rangle $ {\em is determined by the point} $x=x(g)$ {\em of the 
factor space} $G/H$ {\em corresponding to the element} $g$:
\[ 
|\psi _g\rangle =e^{i\alpha }\,|x\rangle ,\qquad |\psi _0\rangle =|0\rangle . 
\]

Let us consider some general properties of the coherent states. It is easy 
to see that $e^{i\alpha (h_2h_1)}=e^{i\alpha (h_2)}\,e^{i\alpha (h_1)}$, 
i.e. $e^{i\alpha (h)}$ is a one-dimensional unitary represenataion of the 
group $H$.

If this representation is not identity, i.e. if $\alpha (h)\neq 0$, then 
the factor group $A$ of the group $H$ on its commutant $H'$\,\,\footnote{\,\,
Remind that the commutant $H'$ of the group $H$ consists of the elements $h'$ 
of the type of $h'=h_1h_2h_1^{-1}h_2^{-2}$. The commutant is an invariant 
subgroup of the group $H$ and the factor group $H/H'$ is the Abelian group.} 
is not trivial, i.e. it contains elements different from unity and the 
character of the group $A$ determines completely the representation of 
the group $H$.

If $\alpha (h)\equiv 0$, then $H$ is the stationary subgroup of the vector 
$|\psi _0\rangle $ in the usual sense. In the first (second) case the 
representation $T$ of the group $G$ being restricted on the subgroup $H$ has 
to contain the one-dimensional (identity) representation of the group $H$\,
\footnote{\,\,
In many cases a useful information on possible representations $T(g)$ may 
be obtained from the reciprocity theorem of Frobenius, stating that if 
$T_\alpha $ is the representation  of the group $G$ induced by the character 
$e^{i\alpha }$ of the group $H$, then the representation $T$ has to be 
conatined in the decomposition of representation $T_\alpha $ into 
irreducible representation [6].}.

Note that if the subgroup $H$ is connected then the  vector $|\psi _0
\rangle$ is the eigenvector of the infinitesimal operators of the 
representation of the subgroup itself.

Let us consider now the action of the operator $T(g)$ on the state 
$|\psi _0\rangle =|0\rangle $
\be T(g)\,|0\rangle =e^{i\alpha (g)}\,|x(g)\rangle . \ee
Here the function $\alpha (g)$ is determined on the whole group $G$ and at 
$g\in H$ it coincides with the previously considered function $\alpha (h)$\,
\footnote{\,\,Formula (1) defines the mapping $\pi \colon G\to \tilde M$ 
where $\tilde M$ is the fiber bundle the base of which is $M=G/H$ and the 
fiber is a circle.}. Substituting in Eq.(1) $g$ by $gh$ we get
\be \alpha (gh)=\alpha (g)+\alpha (h). \ee

Let us now act by the operator $T(g)$ on an arbirary coherent state 
$|x\rangle $
\be T(g_1)\,|x\rangle =e^{-i\alpha (g)}\,T(g_1)\,T(g)\,|0\rangle =
e^{i\beta (g_1,g)}\,|g_1\cdot x\rangle .\ee
Here $\beta (g_1,g)=\alpha (g_1\cdot g)-\alpha (g)$; $x=x(g)$, $g_1\cdot x=x_1
\in M$ and the element $x_1$ is determined by the action of the group $G$ on 
the homogeneous space $M=G/H$. Note that due to (2), Eq.(3) is correct, 
i.e., the right-hand side of equality depends not on $g$ but only on the 
cosets of $x(g)$: $\beta (g_1,g)=\beta (g_1,x)$.

It can be easily seen that the scalar product of two coherent states 
$|x_1\rangle =|x(g_1)\rangle $ and $|x_2\rangle =|x(g_2)\rangle $ is of 
the form
\be \langle x_1|x_2\rangle =e^{i[\alpha (g_1)-\alpha (g_2)]}\langle 0|\,
T(g_1^{-1}g_2)\,|0\rangle \ee
and due to (2) it is independent of the choice of the representatives $g_1$ 
and $g_2$. But due to unitarity of the representations $T(g)$, $|\langle x_1|
x_2\rangle |< 1$ at $x_1\neq x_2$ and the following equalities take 
place
\be \langle x_1|x_2\rangle =\overline{\langle x_2|x_1\rangle }, \ee
\be \langle g\cdot x_1|g\cdot x_2\rangle =e^{i[\beta (g,x_1)-\beta (g,x_2)]}
\langle x_1|x_2\rangle . \ee

Turning to the problem of completeness, note first of all that the 
completeness of the system follows immediately from the irreducibility of 
the representation $T$. Let it exist the invariant measure $dg$ on the group 
$G$. In many cases it induces the invariant mesure $dx$ on the homogeneous 
space $M=G/H$. Supposing the convergence conditions to be fulfilled let us 
consider the operator
\be B=\int dx\,|x\rangle \,\langle x|. \ee
From definition of $B$, the invariance of measure $dx$ and from formula 
(3) it immediately follows that
\be T(g)\,BT(g)^{-1}=B. \ee
Thus $B$ commutes with all the operators $T(g)$ and so due to irreducibility 
of the representation $T$, the operator $B$ is multiple of the identity 
operator
\be \frac1{d}\,B=I. \ee
To find the constant $d$ let us calculate the average value of the operator 
$B$ in the state $|y\rangle $ ($\langle y|y\rangle =1$)
\be \langle y|B|y\rangle =\int |\langle y|x\rangle |^2dx=\int |\langle 0|x
\rangle |^2dx=d. \ee
Hence it is, in particular, seen that a necessary condition for the existence 
of the operator $B$ is the convergence of the integral (10). In this case, 
which we call the case of the square-integrable system of coherent states, 
an importnat identity holds
\be \frac1{d}\,\int dx\,|x\rangle \,\langle x|=I. \ee

Making use of this one may expand the arbitrary state in coherent states
\be |\psi \rangle =\frac1{d}\,\int dx\,c(x)\,|x\rangle ,\qquad c(x)=
\langle x|\psi \rangle . \ee
Here
\be \langle\psi |\psi \rangle =\frac1{d}\,\int dx\,|c(x)|^2 \ee
and the function $c(x)$ is not arbitrary but it must satisfy the condition
\be c(x)=\frac1{d}\,\int \langle x|y\rangle \,c(y)\,dy. \ee

Thus the kernel $K(x,y)=(1/{d})\,\langle x|y\rangle $ is the reproducing one
\be K(x,z)=\int dy\,K(x,y)\,K(y,z) \ee
and the function $\hat f(x)=\int K(x,y)\,f(y)\,dy$ satisfies Eq.(14) for an 
arbitrarily  chosen function $f(x)$.

It can be also easily seen that between the coherent states there are "linear 
dependences". Indeed, from (12) it follows that
\be |x\rangle =\frac1{d}\,\int \langle y|x\rangle \,|y\rangle \,dy . \ee

It means that the system of coherent states is overcomplete, i.e., it 
contains subsystems of coherent states which are complete systems.

The simplest subsets arise from consideration of the discrete subgroups of 
the group $G$. Let $\Gamma $ be a discrete subgroup of the group $G$ such 
that the volume $V_\Gamma $ of the factor space $M/\Gamma $ is finite. 
Let us consider the subsystem of the coherent states
\be \{|x_l\rangle \}, \qquad x_l=x(\gamma _l),\qquad \gamma _l\in \Gamma . \ee

A question arises concerning the completeness of such subsystems. It would 
be interesting to know whether the following statement is valid: at 
$V_\Gamma > d$ the system of states $\{|x_l\rangle\}$ is not complete 
but at $V_\Gamma <d$ this system is complete and remains complete 
even after eliminating any finite number of states. The most interesting case is 
characterized by $V_\Gamma =d$ (if such a condition can be fulfilled) and 
it requires a separate, more detailed consideration. Note that for the 
simplest nilpotent group this problem has been solved in the paper [7].

Let us now illustrate the concept of generalized coherent states using 
concrete examples.

\section{Case of the Special Nilpotent Group}
\setcounter{equation}{17}

This group appears if one writes the Heisenberg commutation relations in the 
Weyl form [4]\,\footnote{\,\,Its properties are considered in details in 
the paper [8].}. 

Let us first review some well known facts. The Lie algebra of this group is 
isomorphic to the Lie algebra produced by the annihilation operators 
$a_1,\ldots ,a_N$, Hermitian-conjugate creation operators $a_1^+,\ldots ,
a_N^+$ and the identity operator $I$. The commutation relations between 
these operators are of the form
\be [a_i,a_j]=[a_i^+,a_j^+]=[a_i,I]=[a_j^+,I]=0,\qquad [a_i,a_j^+]=
\delta _{ij}I. \ee

A general element of the Lie algebra can be written as 
\be tI+i\left( \bar \alpha a-\alpha a^+\right) , \ee
where $t$ is a real number, $\alpha =(\alpha _1,\ldots ,\alpha _N)$, 
$a=(a_1,\ldots , a_N)$ are $N$-dimensional vectors , $\alpha _i$ and ${\bar 
\alpha }_i$ are complex conjugate to each other. Here and in the following 
we use an abbreviated notation for the scalar product of two such vectors
\[ \bar \alpha a=\sum _{i=1}^N {\bar \alpha }_i a_i,\qquad \alpha a^+=\sum 
_{i=1}^N \alpha _ia_i^+\,. \]
The Lie group $W_N$ is obtained from the Lie algebra by means of the 
exponential mapping. Thus to the element of the algebra (19) corresponds 
the element $g$ of the group which is denoted by $(t,\alpha )$. Then the 
multiplication law in $W_N$ is given by the formula
\be (s,\alpha )(t,\beta )=(s+t+\mbox{Im}\,(\alpha \bar \beta ),\,\alpha 
+\beta ). \ee
The operators of the irreducible unitary representation of the group $W_N$ 
are of the form
\be T(g)=T(t,\alpha )=e^{it}\,D(\alpha ). \ee

From (21) it follows in particular that to the set of elements $G_0=\{ g_k
=(2\pi k,0)\}$ ($k$ is integer) corresponds the identity operator, i.e., 
the representation under consideration is not a faithful one.

Making use of the commutation relations (18) we can easily obtain the 
multiplication law of the operators
\be D(\alpha )\,D(\beta )=e^{i\,\mbox{\small Im}\,(\alpha \bar \beta )}\,
D(\alpha +\beta ).\ee

Let us now take some vector $|\psi _0\rangle $ in the representation space. 
We denote the stationary subgroup of this vector as $H$. Consider two 
different cases.

{\bf I}. Let $|\psi _0\rangle $ be an arbitrary vector of the Hilbert space 
${\cal H}$. 
In this case the subgroup $H$ consists of the elements of type $(t,0)$ and 
the factor space $M=W_N/H$ is the $N$-dimensional complex space ${\bf C}^N$. 
In correspondence with Section 2 of this paper, the system of generalized 
coherent states is the set of vectors where
\be |\alpha \rangle =D(\alpha )\,|\psi _0\rangle . \ee

Note that the usual system of coherent states whose properties have been 
considered in detail for instance in the papers [1--3] corresponds to the 
choice in (23) of the so called "vacuum" vector $|0\rangle $\,\footnote{\,\,
The vacuum vector $|0\rangle $ is determined by the equations $a_i\,
|0\rangle =0$.} as initial vector $|\psi _0\rangle $. However, it can be 
readily shown that the system of generalized coherent states possesses 
the same properties as the usual system of coherent states. 
In particular the main identity
\be 
\frac1{\pi ^N}\int d^{2N}\alpha \,|\alpha \rangle\,\langle\alpha |=I,\qquad 
d^{2N}\alpha =\prod _{i=1}^N d\,\mbox{Re}\,\alpha _i\cdot d\,\mbox{Im}\,
\alpha _i \ee
remains valid for it.

{\bf II}. Let us extend the Hilbert space  ${\cal H}$ up to the space of 
generalized functions ${\cal H}_{-\infty }$\,\footnote{\,\, The definition 
of space ${\cal H}_{-\infty }$ and the consideration of some of its 
properties  is given in paper [8].} and try to find the eigenvector 
$|\theta _0\rangle \in {\cal H}_{-\infty }$ of the operators $T(h)$, 
where $h=(t,\alpha _n)$ is the element of the subgroup $H$, $\alpha _n=
\sum _{j=1}^{2N}n_j\,\omega _j$, $n_j$ are integer numbers. 
The set of vectors $\alpha _n$ form a lattice $L$ in the space ${\bf C}^N$ 
and we suppose the periods $\omega _j$ of this lattice to be really 
linearly independent. It can be easily seen that the commutant $H'$ of the 
group $H$ consists of the elements $h'=(t_n,0)$, where $t_n=2\pi \sum 
_{ij} B_{ij}n_in_j$ and the $2N\times 2N$ matrix $B$ has the form $B_{ij}=
(1/{\pi })\,\mbox{Im}\,(\omega _i{\bar \omega }_j)$. But according to 
Section 2 the operator $T(h')$ must be equal to the identity operator, and 
from this follows that the elements of the matrix $B$ are integer numbers:
\be B_{ij}=\frac1{\pi }\,\mbox{Im}\,(\omega _i{\bar \omega }_j)\equiv 
0\,(\mbox{mod}\,1). \ee
we call such lattice $L$ admissible\,\footnote{\,\,Note that the factor space 
$M=G/H$ is a complex torus and in the case of admissible lattice it can be 
considered  as an Abelian variety.}. The conditions $T(h)\,|\theta _0\rangle =
e^{i\,\alpha (h)}|\theta _0\rangle $ are in this case equivalent to 
the equations
\be 
D(\omega _i)\,|\theta _0\rangle =e^{i\pi \varepsilon _i}\,|\theta _0\rangle 
\,. \ee
The state $|\theta _0\rangle $ is determined by the real numbers 
$\varepsilon _i$, $i=1,\ldots ,2N$. 
Acting with the operator $D(\alpha )$ on $|\theta _0\rangle $ 
we get the system of generalized coherent states
\be  |\theta _\alpha \rangle =D(\alpha )\,|\theta _0\rangle ,\ee
where $\alpha $ runs over the complex torus $M=G/H$. It appears that the 
states $|\theta _\alpha \rangle $ under the choice of a definite realization 
of space ${\cal H}$ coincide in the essence with the theta functions. Some 
properties of the theta functions in frame of such approach are considered 
in the papers [7,8].

\section{Case of the Simplest Compact Group -- $SU(2)$ Group}
\setcounter{equation}{27}

The $SU(2)$ group is the unitary matrix group of the second order with 
unity determinant. It is locally isomorphic to the $SO(3)$ group -- the 
rotation group of three-dimensional space and it is the most investigated 
group among all the non-Abelian Lie groups\,\footnote{\,\,The properties of 
this group are considered in details e.g. in the books [9,10].}.  
Nevertheless, the coherent states for this group as a special system do not 
appear to have been considered so far.

Let us first review some well known facts. The $T^j$ representation of this 
group is determined by the non-negative number $j$, integer or half-integer. 
The  basis vectors $|j,\mu \rangle $ of the space in which the representation 
acts are labelled by a number $\mu $ that takes the $2j+1$ integer (if $j$ is 
integer) or half-integer (if $j$ is half-integer) values from $-j$ up to $j$. 
The vectors $|j,\mu \rangle $ satisfy the condition
\be T^j(h)\,|j,\mu \rangle =e^{i\mu \varphi }\,|j,\mu \rangle ,\qquad 
h=\left( \begin{array}{ll} e^{i\varphi /2} &0\\
0 & e^{-i\varphi /2}\end{array} \right) . \ee
Here $h$ is the element of the rotation subgroup $H$ around the axis $x_3$. 
Note that the vector $|j,\mu \rangle $ is the eigenvector of the 
infinitesimal 
operator $J_3$ of the representation $T^j$ corresponding to the subgroup $H$
\be J_3|j,\mu \rangle =\mu |j,\mu \rangle . \ee
The factor space $G/H$ is the two-dimensional sphere $S^2$, the point of this 
sphere is determined by the unit vector ${\bf n}$, ${\bf n}^2=1$. Let 
$g({\bf n})$ be the element of the group $G$ which transforms the vector 
${\bf n}_0=(0,0,1)$ into the vector ${\bf n}=(\sin \theta \,\cos \varphi ,\,\,
\sin \theta \,\sin \varphi ,\,\,\cos \theta )$. As $g({\bf n})$ we may 
choose, for example, the element $g_\varphi ^3\,g_\theta ^2$ where 
$g_\varphi ^3$ corresponds to rotation around the axis $x_3$ by the angle 
$\varphi $ and $g_\theta ^2$ to rotation around the axis $x_2$ by the angle 
$\theta $. 
We thus come to the system of coherent states $\{|\mu ,{\bf n}\rangle\}$:
\be 
|\mu ,{\bf n}\rangle =T(g({\bf n}))\,|\mu \rangle =T(g_\varphi ^3)\,
T(g_\theta ^2)\,|\mu \rangle . \ee
(Here we have to omit for simplicity index $j$ to abbreviate the notations).

The system of coherent states can be also determined up to the phase factor 
$e^{i\alpha ({\bf n})}$ using the equation
\be (n_iJ_i)\,|\mu ,{\bf n}\rangle =\mu \,|\mu ,{\bf n}\rangle , \ee
where $J_i$ is the infinitesimal operator of the represenatation which 
corresponds to a rotation around the axis $x_i$. Here $\mu $ should be 
considered as a fixed parameter and the vector ${\bf n}$ as a variable 
quantity.

Let us consider the properties of this system. The scalar product of two 
coherent states is generally speaking non-zero and it equals
\be  \langle\mu ,{\bf n}'|\mu ,{\bf n}\rangle =e^{i\phi ({\bf n}', {\bf n})}\,
d^j_{\mu \mu }(\theta )=e^{i\phi }\left( \cos \frac{\theta }2\right) 
^{2|\mu |}\,P^{(0,2|\mu |)}_{j-|\mu |}(\cos \theta), \ee
where $\cos \theta ={\bf n}'{\bf n}$, $d^j_{\mu \nu }(\theta )$ are standard 
matrix elements of the $SU(2)$ group [9,10] and $P_n^{(a,b)}$ are Jacobi 
polynomials. Hence we find that
\be 
d=\int |\langle\mu ,{\bf n}'|\mu ,{\bf n}\rangle |^2\,d{\bf n}=
\frac{4\pi }{2j+1} \ee
and correspondingly
\be 
\frac{2j+1}{4\pi }\,\int d{\bf n}\,|\mu ,{\bf n}\rangle \,\langle \mu ,
{\bf n}|=I. \ee
Especially simple is the system of coherent states for $\mu =j$. Then, for 
instance, we get from Eq.(32)
\be 
|\langle j,{\bf n}'|j,{\bf n}\rangle |^2=\left( \frac {1+{\bf n}'
{\bf n}}2\right) ^{2j}.\ee

Let us also give the expression for the coherent states in the so-called 
$z$-representation [9,10]. In this case the representation $T$ acts in the 
space of polynomials of $2j$ degree and the operator of the representation 
is given by the formula
\begin{eqnarray}
T^j(g)\,f(z) &=& (\beta z+\bar \alpha )^{2j}\,f\left( \frac{\alpha z-
\bar \beta }{\beta z+\bar \alpha }\right) ,\nonumber \\
&&\\
g&=& \left( \begin{array}{cc} \alpha &\beta \\ -\bar \beta & \bar \alpha 
\end{array} \right) ,\qquad |\alpha |^2+|\beta |^2=1.\nonumber 
\end{eqnarray}
The basis vectors $|j,\mu \rangle $ in this representation are of the form
\be 
\langle z|j,\mu \rangle =\sqrt{\frac{(2j)!}{(j+\mu )!\,(j-\mu )!}}\,
z^{j+\mu }. \ee
In this case from (30) and (36) it immediately follows
\be 
\langle z|\mu ,{\bf n}\rangle =\sqrt{\frac{(2j)!}{(j+\mu )!\,(j-\mu )!}}\,
(\beta z+\bar \alpha )^{j-\mu }\,(\alpha z-\bar \beta )^{j+\mu }, \ee
where
\[ \alpha =\cos \frac{\theta }2\,e^{i\varphi /2},\qquad \beta =\sin \frac
{\theta }2\,e^{i\varphi /2}. \]

Let us map the sphere $S^2$ onto the plane of the complex variables $\zeta $ 
using the stereographic projection
\be \zeta =\mbox{ctg}\,\frac{\theta }2\,e^{i\varphi }. \ee
Expression (38) takes now the form
\be
\langle z|\mu ,{\bf n}\rangle = e^{i\phi }\langle z|\mu ,\zeta \rangle,\quad 
\mbox{where}\quad e^{i\phi }=(-1)^{j+\mu }\,e^{-i\mu\varphi }, \ee
\be \langle z|\mu,\zeta \rangle =\sqrt{\frac{(2j)!}{(j+\mu )!\,(j-\mu )!}}\,
(1+{|\zeta |}^2)^{-j}\,(z+\bar \zeta )^{j-\mu }\,(1-\zeta z)^{j+\mu }. \ee
Acting on the function $|\mu ,\zeta \rangle $ by the operator $T(g)$ we get 
$T(g)\,|\mu ,\zeta \rangle =e^{i\phi }|\mu ,g\cdot \zeta \rangle $, where 
\be 
e^{i\phi }=\left( \frac{{\bar \beta }\zeta +\bar \alpha }
{\beta {\bar \zeta }+\alpha }\right) ^\mu ,\qquad g\cdot \zeta =\frac{\alpha 
\zeta -\beta }{{\bar \beta }\zeta +\bar \alpha }\,. \ee

Let us now give the formula for the scalar product of coherent states in this 
representation
\be 
\langle j,\zeta '|j, \zeta \rangle =(1+|\zeta '|^2)^{-j}\,(1+|\zeta |^2)^
{-j}\,(1+{\bar \zeta }'\zeta )^{2j}. \ee

Note that just as the usual coherent states appear naturally in the problem 
of an oscillator which is under the action of an external time-dependent 
force [1,2] the states $|\mu ,{\bf n}\rangle $ appear naturally when one 
considers the problem of the spin motion in a time-dependent magnetic field. 
In this case the variation of the state over time is determined by the 
Schr\"odinger equation
\be 
i\,\frac{\partial }{\partial t}\,|\psi (t)\rangle =-\,{\bf aJ}\,|\psi (t)
\rangle , \ee
where ${\bf a}=\mu {\bf H}$, $\mu $ is the magnetic moment of the particle, 
${\bf H}(t)$ is the magnetic field, ${\bf J}=(J_1,J_2,J_3)$, $J_i$ is 
the operator of infinitesimally small rotations around the axis $x_i$.

It can be easily seen that if at initial time we have the coherent state 
$|\psi (0)\rangle =|{\bf n}_0\rangle $, then at any following time this state 
remains coherent, i.e.,
\be |\psi (t)\rangle =e^{i\alpha (t)}\,|{\bf n}(t)\rangle , \ee
where the vector ${\bf n}(t)$ is determined by the classical equation of 
motion
\be \dot {\bf n}(t)=-\,[{\bf a}(t),{\bf n}(t)].\ee
The coherent states may be also used to describe the density matrix $\rho $ 
of a particle with spin\,\footnote{\,\,The description of the oscillator 
density matrix using the usual coherent states can be found in the papers 
[1--3].}. Namely, the density matrix $\rho $ is completely determined either 
by the function $P({\bf n})$ or $Q({\bf n})$, according to the formulae
\begin{eqnarray}
\rho &=& \int d{\bf n}\,P({\bf n})\,|{\bf n}\rangle \,\langle {\bf n}|,\\
Q({\bf n}) &=& \langle {\bf n}|\rho |{\bf n}\rangle . \end{eqnarray}
Note one more useful identity
\begin{eqnarray}
|{\bf n}\rangle \,\langle {\bf n}| &=& \frac{2j+1}{16\,\pi ^2}\,
\int c({\bf n},g^{-1})\,T(g)\,dg,\\
c({\bf n},g) &=& \langle {\bf n}|\,T(g)\,|{\bf n}\rangle \end{eqnarray}
which follows from the orthogonality of the matrix elements 
$T_{\mu \nu }^j(g)$. 

Note that if in (47) $P({\bf n})$ is expanded in a series of spherical 
functions
\be P({\bf n})=\sum _{l,m}C_{l,m}\,Y_{l,m}({\bf n}) \ee
one gets the expansion of the density matrix
\be \rho =\sum _{l,m}C_{l,m}\,{\hat P}_{l,m} \ee
in operators
\be 
{\hat P}_{l,m}=\int d{\bf n}\,Y_{l,m}({\bf n})\,|\mu, {\bf n}\rangle \,
\langle \mu, {\bf n}|. \ee

Calculating the integral entering (53) we find
\be 
\langle \nu '|{\hat P}_{l,m}|\nu \rangle =\frac{\sqrt{4\pi \,(2l+1)}}{2j+1}
\,(j,\nu ';\,l,m|j,\nu )(j,\mu ;\,l,0|j,\mu ), \ee
where $(j,\nu ';\,l,m|j,\nu )$ is the Clebsch--Gordan coefficient.

In conclusion note that the formulae obtained in this section carry over 
to the case arbitrary compact Lie group. In order to do this it is only 
necessary to replace the group $H$ by the Cartan subgroup and to take into 
consideration that $2j+1$ is the dimension of the representation $T(g)$ and 
$4\pi $ is the volume of the factor space $M=G/H$. 

\section{Case of the Simplest Noncompact Group -- $SU(1,1)$ Group}
\setcounter{equation}{54}

The $SU(1,1)$ group is the group of unimodular matrices that leave the form 
$|z_1|^2-|z_2|^2$ invariant. The element $g$ of this group has the form
\be g=\left( \begin{array}{cc} \alpha &\beta \\ \bar \beta & \bar \alpha 
\end{array} \right ) ,\qquad |\alpha |^2-|\beta |^2=1. \ee

The $SU(1,1)$ group is isomorphic to the $Sp(2,R)$ group (to the group of 
real symplectic second order matrices) and it is locally isomorphic to the 
$SO(2,1)$ group, the group of "rotations" of the three-dimensional 
pseudoeuclidian space. It has several series of unitary irreducible 
representations and, in particular, two discrete series $T^+$ and $T^-$. 
It is sufficient to consider only one of these, e.g. $T^+$, because all the 
results are automatically carried over to the other case.

The representation of the series $T^+$ is characterized by the positive 
integer or half-integer numbers $k$. It may be realized in the space of the 
function ${\cal F}_k$ that are analytic in the unit disk $|\zeta |\langle 1$ 
and satisfy the condition
\be ||f||^2=\int d\mu _k(z)\,|f(z)|^2 <\infty ,\ee
where
\be d\mu _k(z)=\frac{2k-1}\pi \,\left( 1-|z|^2\right) ^{2k-2}\,d^2z\,. \ee
It is not difficult to see that if $f(z)=\sum _{n=0}^\infty c_nz^n$ then 
\be ||f||^2=\sum _{n=0}^\infty \,\frac{\Gamma (n+1)\,\Gamma (2k)}{\Gamma 
(n+2k)}\,|c_n|^2. \ee

The space ${\cal F}_k$ becomes a Hibert space if the scalar product of two 
vectors is defined in it according to the formula
\be \langle f|g\rangle =\int d\mu _k(z)\,\bar f(z)\,g(z). \ee
Now it can be easily checked that the functions
\be 
|n\rangle =f_n(z)=\sqrt{\frac{\Gamma (n+2k)}{\Gamma (n+1)\,\Gamma (2k)}}\,
z^n \ee
form an orthonormal basis in the space ${\cal F}_k$\,\footnote{\,\,Note 
that if $k$ tends to infinity and $z$ to zero so that $kz$=const then as the 
functions $f_n(z)$, and all other quantities go over into corresponding 
quantities for a special nilpotent group (see Section 3).}.

Let us define the action of the operators $T(g)$ in the space ${\cal F}_k$
\be T^k(g)\,f(z)=(\beta z+\bar \alpha )^{-2k}\,f\left( \frac{\alpha z+
\bar \beta }{\beta z+\bar \alpha }\right) .\ee
It can be shown [10--12] that the operators $T^k(g)$ determine the unitary 
irreducible representation of the $SU(1,1)$ group.

Let us choose in ${\cal F}_k$ the vector of the lowest weight $|0\rangle $ 
as the fixed vector $|\psi _0\rangle $ (the corresponding function $f_0(z)
\equiv 1$). 
Acting on it by the operators $T^k(g)$ we get the system of states
\be |g\rangle = T(g)\,|0\rangle =(\beta z+\bar \alpha )^{-2k}. \ee
The expression can be easily transformed to the form
\be 
|g\rangle =e^{i\phi }\,|\zeta \rangle ;\qquad |\zeta \rangle =(1-|\zeta |^2)^k
\,(1-\zeta z)^{-2k},\qquad |\zeta |< 1. \ee
The set $\{|\zeta \rangle \}$ is just the system of coherent states.

Note that the group of matrices of type $h=\left( \begin{array}{ll}
e^{i\phi /2} & 0\\ 0 & e^{-i\phi /2}\end{array} \right) $ is the 
stationary subgroup $H$ of the vector $|0\rangle $ 
and the factor space $G/H$ is the unit disk. We see that 
according  to Section 2, the coherent state $|\zeta \rangle $ is completely 
determined by the point $\zeta $ of the factor space $G/H$. Note that this 
space could be also realized as an upper sheet of the hyperboloid 
$n_0^2-n_1^2-n_2^2=1$.

Expanding the state $|\zeta \rangle $ on the states $|n\rangle $ we get
\be |\zeta \rangle =\left( 1-|\zeta |^2\right) ^k\,\sum _{n=0}^\infty \sqrt
{\frac{\Gamma (n+2k)}{\Gamma (n+1)\,\Gamma (2k)}}\,\zeta ^n\,|n\rangle .\ee
Hence for the scalar product of two coherent states the following formula 
obtains
\be \langle \zeta '|\zeta \rangle =\left( 1-|\zeta '|^2\right)^k\,\left( 
1-|\zeta |^2\right) ^k\left( 1-{\bar \zeta }'\zeta \right) ^{-2k}. \ee
Correspondingly
\be d=\int d\mu (\zeta )\,|\langle 0|\zeta \rangle |^2=\frac{\pi }{2k-1} \ee
and the condition of completeness takes the form
\be \frac{2k-1}\pi \,\int d\mu (\zeta )\,|\zeta \rangle \,\langle \zeta |=I. 
\ee 
Here $d\mu (\zeta )=(d^2\zeta )/{\left( 1-|\zeta |^2\right) ^2}$ is the 
invariant measure on the disk $|\zeta |<1$. 

Let us now consider an arbitrary normalized vector $|\psi \rangle $ belonging 
to the Hilbert space ${\cal H}$. To this vector the function $\langle \zeta |
\psi \rangle $ may be taken into correspondance and if $|\psi \rangle =\sum 
c_n\,|n\rangle $, then
\be \langle \zeta |\psi \rangle =\left( 1-|\zeta |^2\right) ^k\psi 
(\bar \zeta ), \ee
where
\be \psi (\zeta )=\sum _{n=0}^\infty \sqrt{\frac{\Gamma (n+2k)}
{\Gamma (n+1)\,\Gamma (2k)}}\,c_n\,\zeta ^n. \ee
Note that in this case
\be ||\psi ||^2=\langle \psi |\psi \rangle =\int d\mu _k(\zeta )\,
|\psi (\zeta )|^2, \ee
i.e., $\psi (\zeta )\in {\cal F}_k$. The formula (69) establishes the 
isomorphism  between the spaces ${\cal H}$ and ${\cal F}_k$.

Moreover, from the inequality $|\langle \zeta |\psi \rangle |^2\leq 
||\psi ||^2$ follows a restriction on the growth of function $\psi (\zeta )$
\be |\psi (\zeta )|^2\leq \left( 1-|\zeta |^2\right )^{-2k}\,||\psi ||^2. \ee
From (71) we immediately obtain that strong convergence of the sequence 
$|\psi _n\rangle $ implies pointwise convergence $\psi _n(\zeta )$ uniform 
on any compact subset of the plane $\zeta $.

A characteristic feature of spaces of the type ${\cal F}_k$ are the so called 
"reproducing kernels" which play the role of usual $\delta $-functions. 
Such kernel can be found in the usual way. Namely
\be \delta _{z'}(z)=\sum _{n=0}^\infty \overline{f_n(z')}\,f_n(z)=
(1-{\bar z}'\,z)^{-2k}. \ee
At fixed $z'$, $\delta _{z'}(z)$ is a function of $z$ and its norm is equal 
to 
\be ||\delta _{z'}||^2=\left( 1-|z'|^2\right) ^{-2k}. \ee

It can be easily checked by direct calculations that $\delta _{z'}(z)$ is the 
analog of $\delta $-function, i.e., the equality
\be \langle \delta _z|f\rangle =\int d\mu (z')\,{\bar \delta }_z(z')\,
f(z')\equiv f(z),\qquad (f(z)\in {\cal F}_k) \ee
holds.

Up to now we considered the representations of the $SU(1,1)$ group. One can 
however consider also the representations of its universal covering group, 
namely the group $\overline{SU(1,1)}$\,\footnote{\,\, Note that the group 
$\overline{SU(1,1)}$ is a dynamical symmetry group in some model many body 
problems [13].} which as is well known, covers the group $SU(1,1)$ 
an infinite number of times. It can be easily seen that this results only in 
replacing the number $k$, that was previously restricted to non-negative 
integer or half-integer values, by an arbitrary non-negative number.

Note that analogous results obtain also for other semi-simple Lie groups, 
having a discrete series of representations.

In this paper we have briefly considered the simplest systems of coherent 
states. It would be interesting to consider other systems of such states, 
in particular, the systems related to the continuous spectrum.

\medskip
{\small Thanks are due to F.A. Berezin for the discussion of these results 
and F. Calogero for his help in translating of the paper into English.}


\begin{thebibliography}{lll}
\bibitem[1]{Gl} Glauber, R.J.: Phys. Rev. {\bf 130}, 2529 (1963); {\bf 131}, 
2766 (1963).
\bibitem[2]{KS} Klauder, J.R., Sudarshan, E.C.G.: Fundamentals of quantum 
optics. New York: Benjamin 1968.
\bibitem[3]{PP} Perelomov, A.M., Popov, V.S., Zel'dovich, B.Ya.: JETP 
{\bf 55}, 589 (1968); {\bf 57}, 196 (1969); Preprints ITEP No. {\bf 612}, 
{\bf 618} (1968).
\bibitem[4]{CV} Weyl, H.: Gruppentheorie und Quantenmechanik. Leipzig: 
S.Hirzel 1928.
\bibitem[5]{MN} Barut, A.O., Girardello, L.: Commun. Math. Phys. {\bf 21}, 
41 (1971).
\bibitem[6]{CD} Mackey, G.W.: Bull. Am. Math. Soc. {\bf 69}, 628 (1963).
\bibitem[7]{HY} Perelomov, A.M.: Theoret. Math. Phys. {\bf 6}, 156 (1971).
\bibitem[8]{IU} Cartier, P.: Proc. Symp. Pure Math., V.{\bf 9}, 
Algebraic groups and discontinuous subgroups, 
 p.361, Providence, R.I.: Amer. Math. Soc. 1966.
\bibitem[9]{YA} Gelfand, I.M., Minlos, R.A., Shapiro, Z.Ya.: Representations 
of the rotation group and the Lorentz group. Oxford: Pergamon Press 1963.
\bibitem[10]{MQ} Vilenkin, N.Ya.: Special functions and the theory of group 
representations. Providence, R.I.: Amer. Math. Soc. 1968.
\bibitem[11]{LR} Bargmann, V.: Ann. Math. {\bf 48}, 568 (1947).
\bibitem[12]{MN} Bargmann, V.: Comm. Pure Appl. Math. {\bf 14}, 187 (1961).
\bibitem[13]{PP} Perelomov, A.M.: Theoret. Math. Phys. {\bf 6}, 263 (1971).
\end{thebibliography}
\end{document}